\documentclass[11pt]{article}

\newfont{\tenmsb}{msbm10 scaled\magstep1}

\let\ssection=\section
\renewcommand{\section}{\setcounter{equation}{0}\ssection}
\usepackage{graphics,amssymb,amsmath,amsthm,amscd,array}
\usepackage{graphicx}
\usepackage{makeidx}

\newcommand\half{{\scriptstyle{\frac{1}{2}}}}
\newcommand\const{\mathop{\rm const}\nolimits}
\def\parag{\hfil\break} %%%%% paragraph
\def\kikezd{\parag\underbar}

\def\p{{\partial}}
\def\br{\vec{r}}

\def\pib{{\bf\pi}}
\def\sigmab{{\bf\sigma}}

\def\sigmab{{\bf\sigma}}
\def\2{{\frac{1}{2}}}
\def\parag{\hfil\break} %%%%% paragraph
\def\IR{{\bf R}} %%%%% the Reals
\def\smallover#1/#2{\hbox{$\textstyle{#1\over#2}$}} 
\def\2{{\smallover1/2}}

\def\\{\cr\noalign{\medskip}}
\def\parag{\hfil\break} %%%%% paragraph
\def\={\!=\!}
\def\D{{D\mkern-2mu\llap{{\raise+0.5pt\hbox{\big/}}}\mkern+2mu} }
\def\H{{\cal H}}
\def\K{{\cal K}}
\def\boxit#1{
\vbox{\hrule\hbox{\vrule\kern4pt
\vbox{\kern5pt#1\kern5pt}\kern4pt\vrule}\hrule}
} %%%%% boxit (Knuth)

\begin{document}

\setlength{\baselineskip}{16pt}

\title{The Biedenharn Approach
\\
to
\\
Relativistic Coulomb-type Problems}

\author{
P.~A.~Horv\'athy
\\
Laboratoire de Math\'ematiques et de Physique Th\'eorique
\\ 
Universit\'e de Tours.
\\
Parc de Grandmont. 
F-37 200 TOURS (France)\footnote{e-mail~:horvathy@lmpt.univ-tours.fr}.
}

\date{\today}

\maketitle

\begin{abstract}
The approach developped by Biedeharn in the sixties
for the relativistic Coulomb problem
is reviewed and applied to various physical problems.
\end{abstract}

%\noindent \texttt{hep-th/06mmxxx} 

\tableofcontents

%%%%%%%%%%%%%%%%%%%%%%%%%%%%%%%%%%%%%%%%%%%%%%%%%%%%%%%%%%%%%%%%%%
%%%%%%%%%%%%%%%%%%%%%%%%%%%%%%%%%%%%%%%%%%%%%%%%%%%%%%%%%%%%%%%%%%
\section{Introduction}
%%%%%%%%%%%%%%%%%%%%%%%%%%%%%%%%%%%%%%%%%%%%%%%%%%%%%%%%%%%%%%%%%%
%%%%%%%%%%%%%%%%%%%%%%%%%%%%%%%%%%%%%%%%%%%%%%%%%%%%%%%%%%%%%%%%%%

In a paper anticipating
supersymmetric quantum mechanics, \cite{Bied}, Biedenharn proposed a
new approach to the Dirac-Coulomb problem. His idea has been to 
iterate the Dirac equation.  The resulting quadratic
equation, written in a  non-relativistic
Coulomb form,  is readily solved using 
the \lq Biedenharn-Temple' operator $\Gamma$ analogous to 
the angular momentum operator (but with a fractional eigenvalues).  
Then the solutions of the first-order equation can be recovered 
from those of the second-order equation by projection.

In this review we apply the 
approach of Biedeharn to various physical problems.

%%%%%%%%%%%%%%%%%%%%%%%%%%%%%%%%%%%%%%%%%%%%%%%%%%%%%%%%%%
\section{The Dirac approach}
%%%%%%%%%%%%%%%%%%%%%%%%%%%%%%%%%%%%%%%%%%%%%%%%%%%%%%%%%%%%

Let us first summarize the original approach of Dirac in
his classic book \cite{Dirac}. He starts with the 
first-order Hamiltonian
\begin{equation}
\H=-eA_0+\rho_1\vec{\sigma}\cdot\vec{p}+\rho_3m,
\label{DiracHam}
\end{equation}
where the `Dirac' matrices can be chosen as
\begin{equation}
\rho_1=\left(\begin{array}{ll}
&1_2 \cr1_2 &
\end{array}\right),
\qquad
\rho_2=\left(\begin{array}{ll}&-i1_2 \cr i1_2 & 
\end{array}\right),
\qquad
\rho_3=\left(\begin{array}{ll}1_2 &\cr &-1_2
\end{array}\right),
\end{equation}
where $1_2$ is the $2\times2$ unit matrix.

For a spherically symmetric potential, $A_0\=A_0(r)$,
Dirac proposes the
following solution. First, he proves the vector identity
\begin{equation}
\big(\vec{\sigma}\cdot\vec{u}\big)\,\big(\vec{\sigma}\cdot\vec{v}\big)
=(\vec{u}\cdot\vec{v})+i\,\vec{\sigma}\cdot(\vec{u}\times\vec{v}).
\label{ident}
\end{equation}
Then, applying to the orbital angular momentum and  momentum,
$$
\vec{u}\=\vec{\ell}\equiv\vec{x}\times\vec{p}
\qquad\hbox{and}\qquad
\vec{v}\=\vec{p},
$$ 
respectively,  interchanging
$\vec{u}$ and $\vec{v}$, 
he deduces that the two-component operator
\begin{equation}
z\equiv\vec{\sigma}\cdot\vec{\ell}+1
\end{equation}
anticommutes with $\vec{\sigma}\cdot\vec{p}$,
$ %\begin{equation}
\big\{z, \vec{\sigma}\cdot\vec{p}\big\}=0.
$ %\end{equation}
Therefore, the operator
\begin{equation}
\K=\rho_3Z
\qquad\hbox{where}\qquad
Z=\vec{\Sigma}\cdot
\vec{\ell}+1,
\qquad
\vec\Sigma=\left(\begin{array}{ll}
\vec\sigma&\cr&\vec\sigma
\end{array}\right),
\label{HK}
\end{equation}
commutes with all three terms in the Hamiltonian (\ref{DiracHam})
and is hence a constant ot the motion.

Next, applying to 
$\vec{u}=\vec{v}=\vec{\ell}$ allows him to infer, using the identity
\begin{equation}
\vec{\ell}\times\vec{\ell}=i\vec{\ell},
\end{equation}
that
\begin{equation}
Z^2=
\big(\vec{\sigma}\cdot\vec{\ell}+1\big)^2
=\vec{J}^2+\smallover1/4,
\qquad\hbox{where}\qquad
\vec{J}\equiv\vec{\ell}+\2\vec{\Sigma}.
\end{equation}
$\vec{J}$ is here the total angular momentum operator.
The eigenvalues of ${\cal K}$ are therefore half-integers,
\begin{equation}
\kappa=\pm(j+1/2).
\label{HKspectr}
\end{equation}

Further application of the identity (\ref{ident}) with 
$\vec{u}=\vec{x}$ and $\vec{v}=\vec{p}$ shows that
\begin{equation}
\big(\vec{\sigma}\cdot\vec{x}\big)
\big(\vec{\sigma}\cdot\vec{p}\big)
=rp_r+i(z-1),
\end{equation}
where $p_r=-i\partial_r$. Note that $\big[p_r,\K\big]=0$.

At this stage, Dirac introduces a second operator, namely
\begin{equation}
\omega=\rho_1W,
\qquad
W=\left(\begin{array}{ll}w&\cr&w\end{array}\right),
\qquad
w=\vec{\sigma}\cdot{\vec{x}\over r},
\end{equation}
which satisfies the relations
\begin{equation}
\omega^2=W^2=w^2=1,
\qquad
\big[\omega,\K\big]=0.
\end{equation}

Finally, Dirac rewrites the Hamiltonian (\ref{DiracHam}) in the form
\begin{equation}
\H=-eA_0+\omega\big(p_r+i\displaystyle{Z-1\over r}\big)+\rho_3m.
\label{DirZHam}
\end{equation}

In the Coulomb case, $eA_0=\alpha/r$, and the radial form
(\ref{DirZHam}) allows one
to find the spectrum (\ref{Hspectrum})
of the relativistic hydrogen atom \cite{Dirac}.

%%%%%%%%%%%%%%%%%%%%%%%%%%%%%%%%%%%%%%%%%%%
\section{The Biedenharn approach to the Dirac-Coulomb problem.}
%%%%%%%%%%%%%%%%%%%%%%%%%%%%%%%%%%%%%%%%%%%%%%%%%%%%%%%%%%%%%

Biedenharn \cite{Bied} proposes instead to introduce  the projection operators
\begin{equation}
{\cal O}_{\pm}=
i\rho_2 \vec{\sigma}\cdot\vec{p}\pm m 
-\rho_3(E+{\alpha\over r}),
\end{equation}
so that $\H-E\=\rho_3{\cal O}_+$,
and observes that 
$$
(\H-E)\psi=0 
\;\Rightarrow\;
{\cal O}_- {\cal O}_+
\psi=0,
\qquad
{\cal O}_{+}\phi={\cal O}_{+}{\cal O}_{-}\psi
={\cal O}_{-}{\cal O}+{-}\psi=0,
$$
since the ${\cal O}_{\pm}$ commute. The solutions of the
first-order equation  ${\cal O}_+\phi=0$
can be obtained, therefore, from those of the iterated equation
by projection, 
\begin{equation}
\phi={\cal O}_-\psi=0.
\label{BiedProj}
\end{equation}

Then the `Biedenharn (Temple) operator' is defined as
\begin{equation}
\Gamma=-\big(Z+i\alpha\omega\big)
\equiv-\left(\begin{array}{ll}
z&i\alpha w\cr i\alpha w &z\end{array}\right).
\label{HTemple}
\end{equation}
 
$\Gamma$ is conserved for the iterated, but not 
for the first-order equation, and allows us to re-write 
${\cal O}_- {\cal O}_+\psi=0$ in a form reminiscent of
the non-relativistic Coulomb problem,
\begin{equation}
\Bigg[-(\partial_r+{1\over r})^2+
\displaystyle{\Gamma (\Gamma+1)\over r^2}
+\displaystyle{2\alpha E \over r}+m^2-E^2 \Bigg]\psi=0.
\label{Hiterated}
\end{equation}

The operator $\Gamma$ plays here a r\^ole of  the angular momentum.
However, 
\begin{equation}
\Gamma^2=
{\cal K}^2-\alpha^2=\vec{J}^2+\smallover1/4-\alpha^2,
\end{equation}
so that the eigenvalues of $\Gamma$ are
\begin{equation}
\gamma=\pm \sqrt{\kappa^2-\alpha^2}
=\pm\sqrt{(j+1/2)^2-\alpha^2}, 
\qquad \hbox{sign}\ \gamma=
\hbox{sign}\ \kappa.  
\end{equation}
For a $\Gamma$-eigenfunction,
\begin{equation}
\Gamma (\Gamma+1)
=
\ell(\gamma) (\ell(\gamma)+1)
\qquad\hbox{with}\qquad
\ell(\gamma)=\,\mid \gamma \mid 
+\2\big[{\rm sign}(\gamma)-1\big],
\label{HGammaeigenvalue}
\end{equation}
i.e. the  `angular momentum' $\ell(\gamma)$ is {\it irrational}.
The operator $\Gamma$ is hermitian as long as
$\alpha\leq1$, i.e., for nuclei with less then $137$ protons.
\goodbreak

To get explicit formul{\ae}, remember \cite{BjDr} that the angular spinors
\begin{equation}
\chi^{\mu}_{\pm}=
\sqrt{{\mid\kappa \mid+1/2\pm\mu\over2\mid\kappa\mid+1}} 
\ Y^{\mu -1/2}_{j\pm1/2}
\left(\begin{array}{ll}1\cr0\end{array}\right)
\mp\sqrt{{\mid \kappa\mid+1/2 \mp\mu\over2\mid\kappa\mid+1}} 
\ Y^{\mu+1/2}_{j\pm1/2}
\left(\begin{array}{ll}0\cr1\end{array}\right),
\label{Hangularfunct}
\end{equation}
where the $\pm$ refers to the sign of $\kappa$, and the $Y$'s are the spherical harmonics,
 are not only eigenfunctions of $\vec{J}^2$ and of $J_3$ with
eigenvalues $j(j+1)$, \ $(j=1/2, 3/2,\cdots)$ and $\mu=-j, \ldots,j$
respectively, but also satisfy the crucial relations 
\begin{equation}
z\ \chi^{\mu}_{\pm}
=
\pm\mid\kappa\mid\ \chi^{\mu}_{\pm} 
\qquad\hbox{and}\qquad
w\ \chi^{\mu}_{\pm}=\chi^{\mu}_{\mp}.
\end{equation}

Put
\begin{equation}
\Xi_+=\left(\begin{array}{ll}\chi^{\mu}_+\cr0\end{array}\right),
\qquad
\Xi_-=\left(\begin{array}{ll}0\cr\chi^{\mu}_-\end{array}\right),
\qquad
\Upsilon_+=\left(\begin{array}{ll}0\cr\chi^{\mu}_+\end{array}\right),  
\qquad
\Upsilon_-=\left(\begin{array}{ll}\chi^{\mu}_-\cr 0\end{array}\right).
\label{XiUps}
\end{equation}
Then the
\begin{equation}
\begin{array}{ll}
\Phi_+=-i\alpha \Xi_++(\mid\kappa\mid-\mid\gamma\mid)\Xi_-,
\qquad\hfill
&\Phi_-= (\mid\kappa\mid-\mid\gamma\mid)\Xi_++i\alpha\Xi_- ,
\\\cr
\varphi_+
=-i\alpha\Upsilon_++(\mid\kappa\mid-\mid\gamma\mid)\Upsilon_-, 
\qquad\hfill
&\varphi_-
=(\mid\kappa\mid-\mid\gamma\mid)\ \Upsilon_++ i\alpha\Upsilon_-
\end{array}
\label{HGammaeigenfunc}
\end{equation}
are eigenfunctions of $\Gamma$ with eigenvalues 
$\pm\mid\gamma\mid $,
\begin{equation}\Gamma\Phi_{\pm}=\pm \mid\gamma\mid\Phi_{\pm}
\qquad
\Gamma\varphi_{\pm}=\pm \mid\gamma\mid\varphi_{\pm}.
\label{HGammaeigenval}
\end{equation}

\begin{figure}[htbp]
\centering
\includegraphics[scale=.3]{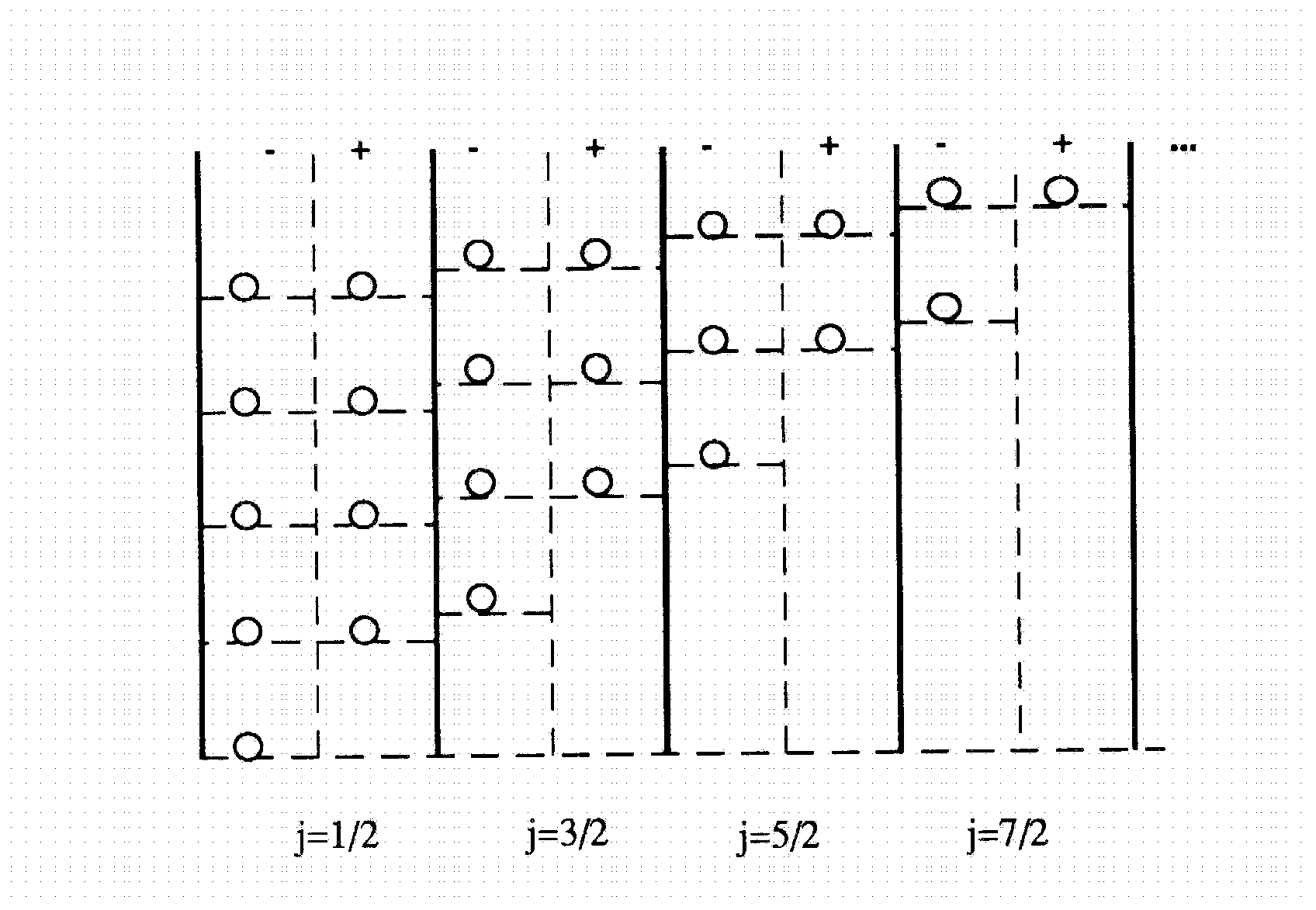}
\caption{\it The spectrum of a Dirac electron in the field of an H-atom. The
$\pm$ signs refer to the sign of $\gamma$. 
In different $j$-sectors
the energy levels are shifted by the fine structure.}
\label{Hatom}
\end{figure}

Then, setting $\psi_{\pm}=u_{\pm} \Phi_{\pm}$,
the iterated equation takes indeed a non-relativistic
Coulomb form with irrational angular momentum $\ell(\gamma)$,
\begin{equation}
\Big[-(\partial _r+{1\over r})^2+ 
{\ell(\gamma)(\ell(\gamma)+1)\over r^2}
+{2\alpha E \over r}+m^2-E^2 \Big]\ u_{\pm}=0,  
\label{HCoulomb}
\end{equation}
whose solutions are the well-known Coulomb eigenfunctions
\begin{equation}
u_{\pm}(r)\propto\ r^{\ell(\gamma)}e^{ikr}\
F\Big(\ell(\gamma)+1-i{\alpha E \over k}, 2\ell(\gamma)+2,-2ikr\Big),
\label{Hstates}
\end{equation}
where $k=\sqrt{E^2-m^2}$ and $F$ denotes the confluent hypergeometric
function. The energy levels are obtained from  
the poles of $F$,
$$
\ell(\gamma)+1-i\alpha E/k=-n,\qquad 
n=0,1, 2,\ldots,
$$
yielding the familiar spectrum shown on FIG. \ref{Hatom},
\begin{equation}
E_p=
m \ \sqrt{1-{\alpha^2\over {p^2+\alpha^2}}}, 
\quad 
p=\ell(\gamma)+1+n=\mid \gamma\mid 
+{1\over2}\hbox{sign}\ \gamma+{1\over 2}+n, 
\quad
n=0, 1, \ldots
\label{Hspectrum}
\end{equation}
%$n=0, 1, \ldots$. 
Since $\gamma$ and thus 
$\ell$ are irrational, $\ell+n=\ell'+n'$ is only  possible for $\gamma'=
\pm\gamma$ so different $j$-sectors yield different $E$-values.
For each fixed $j$, the same energy is obtained in the $\gamma>0$
sector for $n-1$ as in the $\gamma<0$ sector for $n$. 
These energy levels are
hence doubly degenerate. In the $\gamma<0$ sector the $n=0$ 
state is unpaired: {\it each} $j$ sector admits a ground-state.
\begin{equation}
u_0^j\ \propto\  r^{\mid \gamma \mid -1}e^{-\alpha mr/(j+1/2)}
\qquad\hbox{with energy}\qquad 
E_0^j= 
 m\ \sqrt{1-{\alpha^2\over(j+1/2)^2}}.
 \label{Hgroundstate}
\end{equation}

Observe that eqn.
(\ref{Hgroundstate}) is consistent with (\ref{Hstates}) due to $F(a,a,z)=e^z$.
\goodbreak

%%%%%%%%%%%%%%%%%%%%%%%%%%%%%%
\section{Charged Dirac particle in a monopole field}\label{CM}
%%%%%%%%%%%%%%%%%%%%%%%%%%%%%%

A Dirac particle in the field of a Dirac monopole,
\begin{equation}
\vec{B}=-g{\vec{r}\over r^3},
\end{equation}
can be treated along the same lines \cite{BM}. The Hamiltonian is now 
\begin{equation}
\H=-{\alpha \over r}+\rho_1\vec{\sigma}\cdot\vec{\pi}+\rho_3m,
\qquad
\vec\pi=\vec{p}-e\vec{A},
\end{equation}
where $\vec{A}$ is the vector potential of a Dirac monopole,
$\vec\nabla\times\vec{A}=\vec{B}$.
Introducing again the projection operators 
\begin{equation}
{\cal O}_{\pm}
=
i\rho_2\,\vec{\sigma}\cdot\vec{\pi}
\pm m-\rho_3
(E+{\alpha \over r}),
\end{equation}
the solutions of the
first-order equation  can again be obtained from that of the 
iterated equation by projecting, cf. (\ref{BiedProj}). 
Dirac's operator,  
\begin{equation}
{\cal K}=-\rho_3 \big(\vec{\Sigma}\cdot\vec{\ell}+1\big),
\label{DiracK}
\end{equation} 
is {\it formally} the same as in (\ref{HK}), except for the replacements
\begin{equation}
\vec{p}\to\vec\pi,
\qquad\Rightarrow\qquad
\vec\ell=\vec{r}\times \vec{\pi}.
\end{equation}

Note that $\vec{\ell}$ is now only part of the orbital angular momentum,
$$
\vec{L}=\vec{\ell}-q\frac{\vec{r}}{r},
$$
where $q=eg$.
The novelty is that, unlike in (\ref{HKspectr}), the eigenvalues of $\K$ 
became now irrational,
\begin{equation}
\kappa=\sqrt{(j+1/2)^2-q^2}.
\label{MKspect}
\end{equation}

The iterated equation reads again as (\ref{Hiterated}), with the Biedenharn operator 
$
\Gamma=-(Z+i\alpha\omega)
$
cf. (\ref{HTemple}). 
The square of $\Gamma$ is now 
\begin{equation}
\Gamma^2={\cal K}^2-\alpha^2
=\vec{J}^2+\smallover1/4-q^2-\alpha^2,
\label{MGammasq}
\end{equation}
where
\begin{equation}
\vec{J}=\vec{L}+\2\vec{\Sigma}=\vec{\ell}-q{\vec{r}\over r}+\2\vec{\Sigma}
\label{MTotangmom}
\end{equation}
is the total angular momentum.
The eigenvalues of $\Gamma$ are, therefore, `even more irrational', since
the monopole-charge term $q^2$ and the Coulomb-charge term $\alpha^2$ 
are both subtracted~: 
\begin{equation}
\gamma=
\pm\sqrt{\kappa^2-\alpha^2}=\pm\sqrt{(j+1/2)^2-q^2-\alpha^2}, 
\qquad 
\hbox{sign}\ \gamma=\hbox{sign}\ \kappa.
\label{HMGAMMA}
\end{equation}

Observe that this yields
now an imaginary $\gamma$ for the lowest angular momentum
$j=q-1/2$ sector for any positive $\alpha$,
and the situation is worsened when $\alpha$ is increased.
These cases should be discarded.

Let us assume that $\alpha$
is small, typically a few times 1/137 so that $\gamma$ is real
except for the lowest angular momentum sector.
Assuming $j \geq q+1/2$,
consider those angular 2-spinors $\chi_{\pm}$ in (\ref{Hangularfunct}), i.e.
\begin{equation}
\chi^{\mu}_{\pm}=
\sqrt{{\mid\kappa \mid+1/2\pm\mu\over2\mid\kappa\mid+1}} 
\ Y^{\mu -1/2}_{j\pm1/2}
\left(\begin{array}{ll}1\cr0\end{array}\right)
\mp\sqrt{{\mid \kappa\mid+1/2 \mp\mu\over2\mid\kappa\mid+1}} 
\ Y^{\mu+1/2}_{j\pm1/2}
\left(\begin{array}{ll}0\cr1\end{array}\right),
\label{Mpangfunct}
\end{equation}
but with the $Y$'s being now replaced  by the 
`Wu-Yang' {\it monopole harmonics} \cite{WY}.
These spinors are eigenfunctions of
$\vec{J}\strut^2$ and $J_3$ with eingenvalues 
$j=q-\2, q+\2, \dots$ and $-j\leq\mu\leq j$, respectively.
Then the $\Phi_\pm$ and  $\varphi_\pm$ in Eq. (\ref{HGammaeigenfunc}) are eigenfunctions
of $\Gamma$ with eigenvalues $\pm|\gamma|$, cf. (\ref{HGammaeigenval}). 
For the two signs  
\begin{equation}
\Gamma(\Gamma+1)
=
\ell(\gamma)(\ell(\gamma)+1) 
\qquad\hbox{with}\quad
\ell(\gamma)=
\mid \gamma\mid+\2[{\rm sign}(\gamma)-1)] 
\end{equation}
cf. (\ref{HGammaeigenvalue}). Setting 
$\psi_{\pm}=u_{\pm} \Phi_{\pm}$
(and $\psi_{\pm}=u_{\pm} \varphi_{\pm}$, respectively),
the iterated Dirac equation ${\cal O}_- {\cal O}_+$  reduces to
the non-relativistic Coulomb form       
(\ref{HCoulomb})  with  solutions as in (\ref{Hstates}) and energy levels (\ref{Hspectrum}).
The only difference is in the value of $\gamma$. 

The ground-states of the $j\=\const$ sector are  
\begin{equation} 
u_0^j
\ \propto\  
r^{\mid\gamma\mid-1}
e^{-\alpha m r/\sqrt{\gamma^2+\alpha^2}}
\quad\hbox{with energy}\quad
E_0^{(j)}
= 
m\ \sqrt{1-{\alpha^2\over\gamma^2+\alpha^2}}.
\end{equation}

\begin{figure}[htbp]
\centering
\includegraphics[scale=.3]{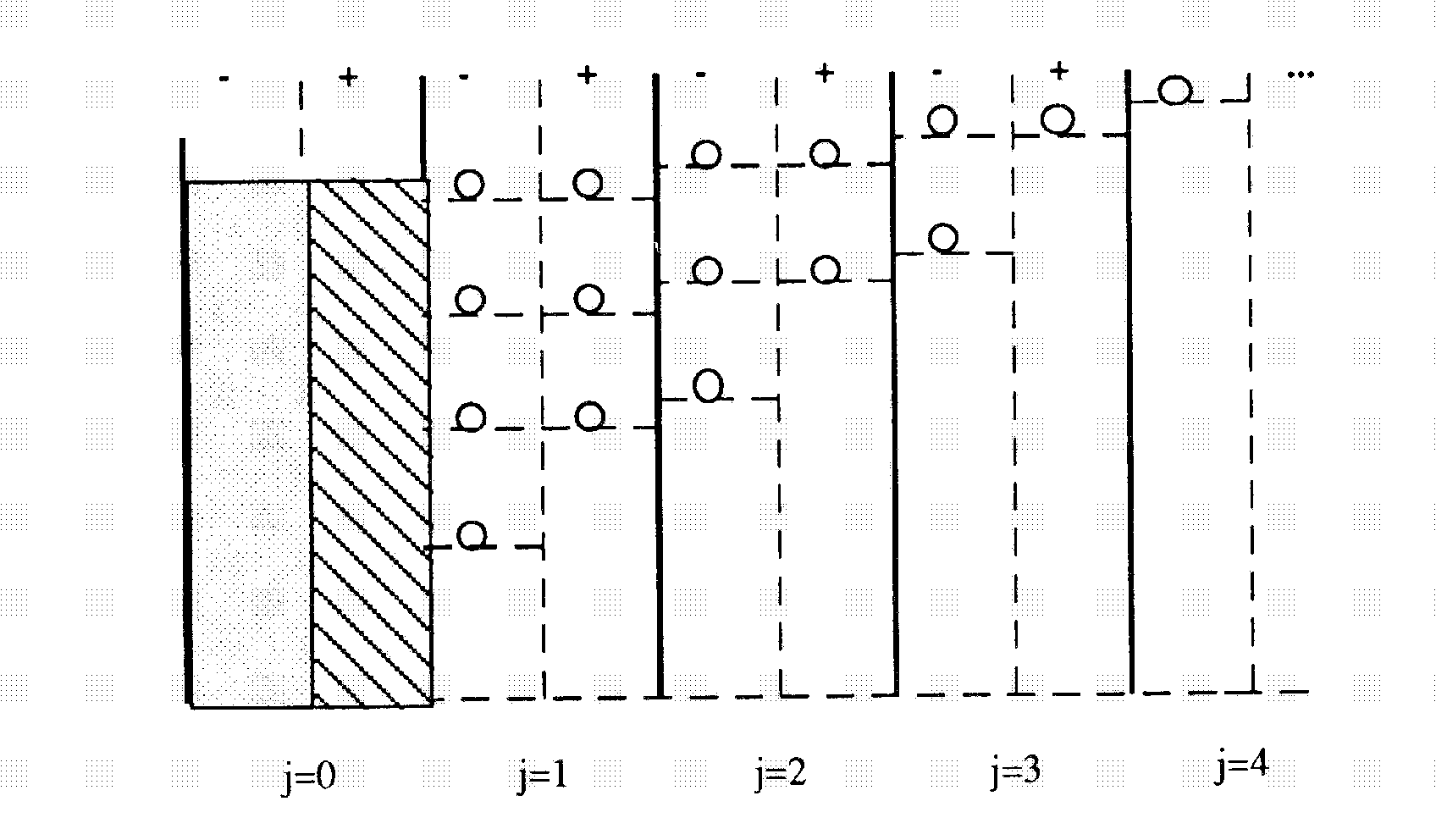}
\caption{\it The bound-state spectrum of a Dirac particle in a charged monopole field [for $q=1/2$]. 
 The $\pm$ refers to the sign of the Biedenharn 
operator  $\Gamma$. 
In each $j=const$. sector the energy levels are doubly
degenerate except for a lowest-energy ground state, which occurs in the
$\gamma<0$ sector. Different $j$-sectors are shifted by a modified fine
structure. For $j=0$ there are no $\gamma>0$ states, and $\Gamma$
is not hermitian. This critical case $j=0$, $\gamma<0$ is not discussed
here.}
\label{Dmon}
\end{figure}

The spectrum is shown on FIG.\ref{Dmon}.

\goodbreak

For $q=0$ (no monopole) we plainly
recover Biedernharn's results in \cite{Bied} on the
Dirac-Coulomb problem.

For $\alpha=0$ (no Coulomb potential) one has a pure Dirac monopole
\cite{Band}. The Biedenharn 
operator $\Gamma$ reduces to $-Z$. No further diagonalization in
$\rho$-space is thus necessary. Since $[Z, \rho_3] = 0$, $\rho_3$ is 
 now conserved for the iterated equation (but
not for the first-order equation). The iterated equation splits therefore into
two (identical) Pauli equations, and we can work with 2-spinors.
  
For $j\geq q+1/2$, the angular eigenfunctions of $\Gamma=-Z$  
those $\Xi$'s in (\ref{XiUps}). 
\footnote{the $\Xi_{\pm}$'s are proportional to those $\xi^i$'s ($i=1,2$) 
in eqns. (11) and (19) of Kazama, Yang and Goldhaber \cite{KYG}.
Their $\phi^i$'s are just our $\varphi_{\mp}$'s in (\ref{HGammaeigenfunc}).}. 

For $j \geq q+1/2$ there are no
bound states. The hypergeometric
function reduces to a Bessel function and the
radial eigenfunction becomes 
\begin{equation}
u_{\pm}
\ \propto\ 
{1\over\sqrt{kr}} J_{\mid\kappa\mid\pm1/2}.
\end{equation}
the same as eqn. \# (37) in \cite{DV84}.

The $j=q-1/2$ case should {\it not} be discarded:
the eigenvalue of $Z\equiv \Gamma$ only vanishes,
rather then becoming imaginary.
The problem requires, nevertheless,  special treatment. %£
The Dirac Hamiltonian is indeed not self-adjoint \cite{Callias}
but admits a 1-parameter family of
self-adjoint extensions, corresponding to different boundary conditions at $r=0$. These yield different physics. 
The one constructed by  Callias \cite{Callias} has further
significance for the theta-angle in QCD.
Kazama et al. \cite{KYG,KY} suggest to cure the non-self-adjointnsess problem by
adding an infinitesimal extra magnetic moment. 
 For further discussion and details
the reader is invited to consult the literature 
 \cite{Callias,KYG,KY,SelfAdj}.
%%%% INSERT UP TO HERE

%%%%%%%%%%%%%%%%%%%%%%%%%%%%
\section{Dyons}\label{dyon}
%%%%%%%%%%%%%%%%%%%%%%%%%%%

Let us consider a massless Dirac particle in the long-distance field 
of a (self-dual)
Bogomolny-Prasad-Sommerfield monopole \cite{FHO,Feher,Bloore},
\begin{equation}
e\vec{B}=-q{\vec{r}\over r^3}
\qquad\hbox{and}\qquad
\Phi=q\left(1-{1\over r}\right).
\end{equation}
%$q=eg$. 
Identifying $\Phi$ with the fourth
component of a gauge field we get a static, self-dual 
Abelian gauge field in four euclidean dimensions
\begin{equation}
{\bf A}=q{\bf A}_D,\qquad
A_4=q\left(1-{1\over r}\right),
\label{dyongaugefield}
\end{equation}
where ${\bf A}_D$ denotes the vector potential of a Dirac
monopole of unit strength.
The associated Dirac Hamiltonian is therefore \cite{FHO,Bloore}
\begin{equation}
\D=\rho_1(\vec{\sigma}\cdot\vec\pi)-\rho_2\Phi\,
=
\left(\begin{array}{ll} &Q^{\dagger}\cr Q &
\end{array}\right)
=\,\left(\begin{array}{ll}&{\bf\sigma}\cdot\vec{\pi}-i\Phi\cr
\vec{\sigma}\cdot\vec{\pi}+i\Phi&\end{array}\right).
\label{DyonDirac}
\end{equation}

In contrast to the
Coulomb case, the scalar term $\rho_2\Phi$ is now {\it off-diagonal}, because
it comes from the fourth, euclidean, direction, rather than from the 
time coordinate. 
%Below we find the spectrum of $\D$.

The total angular momentum, $\vec{J}$
in Eq. (\ref{MTotangmom}) is conserved.
Using the  notations and formul{\ae} introduced for the charged monopole,
we observe that the counterpart of Dirac's operator (\ref{DiracK}),
\begin{equation}
{\cal K}=-\rho_2Z=\left(\begin{array}{ll}&iz\cr-iz&\end{array}\right),
\label{dyonK}
\end{equation}
commutes with $\D$ and
\begin{equation}
{\cal K}^2=z^2=\vec{J}^2+{1\over4}-q^2,
\end{equation}
so that z (and hence $Z$ and $\cal K$) have irrational eigenvalues,  
\begin{equation}
\kappa=\sqrt{(j+1/2)^2-q^2}, 
\label{dyonKspect}
\end{equation}
cf. Eq. (\ref{MKspect}).
Since $j\geq q-1/2$, 
${\cal K}$ is hermitian, but for $j=q-1/2$ its eigenvalue $\kappa$
vanishes and thus ${\cal K}$ is not invertible.
 
The Dirac operator (\ref{DyonDirac}) is, as in any even dimensional space,
chiral-supersymmetric~: 
$\lbrace Q, Q^{\dagger}\rbrace$ 
is a SUSY Hamiltonian and the SUSY sectors are the $\pm1$
eigenspaces of the chirality operator $\rho_3$. 
The supercharges $Q$ and
$Q^{\dagger}$ can be written as 
\begin{eqnarray}
Q&=-iw\Big(\partial_r+\displaystyle{1\over r}-
\displaystyle{z+qw\over r}+qw\Big)=
-i\Big(\partial_r+\displaystyle{1\over r}+
\displaystyle{z-qw\over r}+qw\Big)w,
\\[6pt]
Q^{\dagger}&=iw\Big(-(\partial_r+\displaystyle{1\over r})+
\displaystyle{z-qw\over r}+qw\Big)=
i\Big(-(\partial_r+\displaystyle{1\over r})-
\displaystyle{z+qw\over r}+qw\Big)w.
\end{eqnarray}

The square of (\ref{DyonDirac}) is 
\begin{eqnarray}
\D^2=\left(\begin{array}{ll}H_0,&\\ &H_1
\end{array}\right)=
\left(\begin{array}{ll}Q^{\dagger}Q&\cr&QQ^{\dagger}\end{array}\right),
\end{eqnarray}
where
\begin{equation}
H_0=\left[\pi^2+q^2\left(1-\frac{1}{r}\right)^2\right]1_2
\qquad\hbox{and}\qquad
H_1=H_0-2\frac{\sigmab\cdot\br}{r^3}.
\end{equation}  

In the `lower' (i.e. $\rho_3 = -1$) sector,
the gyromagnetic ratio is $g = 0$, and $H_0$ can be viewed 
as describing two, uncoupled, spin $0$ particles in the combined
field of a Dirac monopole, of a Coulomb potential and of an inverse-square
potential. This system has been solved many years ago;
it has a Coulomb-type spectrum, whose degeneracy is explained by
its `accidental' ${\rm o}(4)$ symmetry \cite{MCZ}.
In the `upper'' (i.e. $\rho_3 = 1$) sector $g=4$;
$H_1$ is the Hamiltonian 
of D'Hoker and Vinet in Ref. \cite{DVdyon}. 

In terms of $Z$ and $w$, $\D^2$ is also
\begin{eqnarray}
\D^2=
-(\partial _r+\displaystyle{1\over r})^2-\displaystyle{2q^2\over r}
+q^2+\displaystyle{Z^2+q^2\over r^2}-\displaystyle{1\over r^2}
\left(\begin{array}{ll}z+qw&\cr
 &z-qw\end{array}\right).
 \label{dyonDiracsq}
\end{eqnarray}
The Biedenharn operator \cite{Bloore}  
\begin{equation}
\Gamma
=-(Z+q\rho_3W)
\equiv
-\left(\begin{array}{ll}z+qw&\cr&z-qw\end{array}\right),
\label{dyonGamma}
\end{equation}
does not commute with $\D$, but it
commutes with $\D^2$; it is thus conserved for the quadratic
dynamics $H_0$ and $H_1$  [but not for the Dirac Hamiltonian $\D$].
In terms of $\Gamma$, $\D^2$ becomes 
\begin{equation}
\D^2=-(\partial_r+{1\over r})^2+ 
{\Gamma(\Gamma+1)\over r^2}-{2q^2\over r}+q^2.
\end{equation}
Now 
\begin{equation}
\Gamma^2=z^2+q^2=\vec{J}^2+{1\over 4},
\end{equation}
because, unlike in (\ref{MGammasq}), 
the $q^2$ comes with a positive sign.
The eigenvalues of $\Gamma$ are, therefore, {\it (half)integers}, 
\begin{equation}
\gamma=\pm(j+1/2), 
\qquad
\hbox{sign}\ \gamma=\hbox{sign}\ \kappa.
\label{dyonGammaspect}
\end{equation}
Hence, for a $\Gamma$-eigenfunction,
\begin{equation}
\Gamma(\Gamma+1)=L(\gamma)(L(\gamma)+1)
\qquad\hbox{where}\qquad
L(\gamma)=j\pm\2.
\label{dyonGGplus}
\end{equation}
(The sign is plus or minus depending on the sign of $\gamma$). 
$L(\gamma)$ is now a (half)integer. Using the notations  
$x=z-qw$ and $y=z+qw$,  the supercharges are written as 
\begin{eqnarray}
Q&=-iw\Big(\partial_r+\displaystyle{1\over r}-
\displaystyle{y\over r}+qw\Big)=  
-i\Big(\partial_r+\displaystyle{1\over r}+
\displaystyle{x\over r}+qw\Big)w, 
\\[6pt]
Q^{\dagger}&=iw\Big(-(\partial_r+\displaystyle{1\over r})+
\displaystyle{x\over r}+qw\Big)
=
i\Big(-(\partial_r+\displaystyle{1\over r})-
\displaystyle{y\over r}qw\Big).
\end{eqnarray}

Note that one can also write
\begin{equation}
\Gamma  =-\left(\begin{array}{cc}
\vec{\sigma}\cdot\vec{L}+1+2qw&0
\\[6pt]
0&\vec{\sigma}\cdot\vec{L}
 +1\end{array}\right)
 =
 \left(\begin{array}{ll}-y&0
 \\[6pt]
 0&-x\end{array}\right). 
\end{equation} 

$x$ and $y$ are self-adjoint,  
$x=x^{\dagger}$, $y=y^{\dagger}$, $w=w^{\dagger}$. 

To find an explicit solution, we construct, cf. (\ref{Mpangfunct}),
 angular 2-spinors
$\varphi_{\pm}^{\mu}$ and  $\Phi_{\pm}^{\mu}$,
which are both eigenfunctions of 
$\vec{J}^2$ and $J_3$ with eigenvalues $j(j+1)$ and $\mu$, 
and which diagonalize
the operators $x$ and $y$:
\begin{equation}
x \varphi_{\pm}^{\mu}=\mp\mid\gamma\mid \varphi^{\mu}_{\pm}
\qquad\hbox{and}\qquad
y \Phi_{\pm}^{\mu}=\mp\mid\gamma\mid\Phi^{\mu}_{\pm}.
\end{equation}

In the `lower' sector,  the coefficient of the
$r^{-2}$ term here is the square of the orbital angular momentum, 
\begin{equation} 
x(x-1)=\vec{L}^2=
L(\gamma)(L(\gamma)+1), 
\end{equation}
so that $L(\gamma)$ is just the orbital angular quantum number.
Due to the addition theorem of the angular momentum, if $j \geqÊ
q+1/2$, $L(\gamma)=j\pm1/2$, but for $j=q-1/2$
the only allowed value of $L(\gamma)$ is $L(\gamma)=j+1/2$.

For $j\geq q+1/2$ consider, therefore, 
\begin{equation}
\varphi_{\pm}^{\mu}=\sqrt{{L(\gamma)+1/2\pm\mu\over2L(\gamma)+1}}\, 
Y^{\mu-1/2}_{L(\gamma)}
\left(\begin{array}{ll}1\cr 0\end{array}\right)
\pm 
\sqrt{{L(\gamma)+1/2\mp\mu\over 2L(\gamma)+1}}\, 
Y^{\mu +1/2}_{L({\gamma})}
\left(\begin{array}{ll}0\cr1\end{array}\right),
\label{dyonangfunct}
\end{equation} 
where the $Y$'s are again the Wu-Yang \cite{WY}
monopole harmonics, and the
sign $\pm$ refers to the sign of $\gamma$. The $\varphi$'s 
satisfy\footnote{the superscript $\mu$ is dropped for the sake of simplicity.} 
\begin{eqnarray}
\vec{J}^2 \varphi_{\pm} = j(j+1)\varphi_{\pm}, 
\\[6pt]
J_3 \varphi_{\pm}=\mu\varphi_{\pm}, 
\quad \mu = -j, \cdots,j,
\\[6pt]
\vec{L}^2 \varphi_{\pm}=L(\gamma) 
(L(\gamma)+1)\varphi_{\pm}.
\end{eqnarray}
Since 
$
\vec{L}\cdot\vec{\sigma}=\vec{J}^2-\vec{L}^2-3/4,
$
we have 
\begin{equation}
x\ \varphi_{\pm}=\Big( \vec{L}\cdot\vec{\sigma}+1\Big)\ \varphi_{\pm}=
\mp\mid\gamma\mid\varphi_{\pm},
\end{equation}
as wanted.

For $j=q-1/2$ no $\varphi_-$ (i.e. no $L(\gamma)=q-1$) state 
is available, but
eqn. (\ref{dyonangfunct}) still yields $2(q-\2)+1=2q$ $\varphi_+^0$s
 with $L(\gamma)=q$, namely
\begin{equation}
\left(\varphi_+^0\right)^{\mu}
=\sqrt{{q+1/2+
\mu\over2q+1}}\,
Y^{\mu-1/2}_q
\left(\begin{array}{ll}1\cr0\end{array}\right) 
+
\sqrt{{q+1/2 +
\mu\over2q+1}}\, 
Y^{\mu+1/2}_q
\left(\begin{array}{ll}0\cr1\end{array}\right),
\label{dyonlowestangfunct}
\end{equation} 
where $\mu=-(q-1/2),\ldots,(q-1/2)$.
They are eigenstates of $x$ with eigenvalue $-q$.

The $y$-eigenspinors $\Phi$ of the `upper' (i.e. $\rho_3 = 1$) sector are
constructed indirectly. Assume first that one can find angular spinors
$\chi_{\pm}$ which diagonalize $z=\vec\sigma\cdot\vec\ell+1$,  
\begin{equation}
z\ \chi_\pm
=
\pm\mid\kappa\mid
\chi_{\pm},
\label{dyonz}
\end{equation}
and also satisfy 
\begin{eqnarray}
\vec{J}^2\chi_{\pm}^{\mu}=j(j+1)\, \chi_{\pm}^{\mu},
\qquad
&j=q-1/2, q+1/2,\cdots
\\[6pt] 
J_3\chi_{\pm}^{\mu}=\mu\ \chi_{\pm}^{\mu}, 
\qquad
&\mu=-j,\cdots, j
\\[6pt]
w\ \chi_{\pm}^{\mu}=\chi_{\mp}^{\mu}.
\end{eqnarray}

In the subspace spanned by the $\chi_{\pm}$'s, $x=z-qw$ and 
$y=z+qw$ have the remarkably symmetric matrix representations
\begin{equation}
\Big[x\Big]
=
\left(\begin{array}{ll}\mid\kappa\mid&-q\cr -q&-\mid\kappa\mid\end{array}\right)
\qquad\hbox{and}\qquad
\Big[y\Big]
=
\left(\begin{array}{ll}\mid\kappa\mid&q \cr q&-\mid\kappa\mid\end{array}\right).
\end{equation}

The eigenvectors $\varphi_{\pm}$ and $\Phi_{\pm}$ of $x$ and $y$ with
eigenvalues $\pm \mid \gamma \mid$ are thus
\begin{equation}
\begin{array}{ll}
\varphi_+=(\mid\kappa\mid+\mid\gamma\mid)\chi_+-q\chi_-, 
&\varphi_-=
q\chi_++(\mid\kappa\mid+\mid\gamma\mid)\chi_-,
\\[10pt]
\Phi_+=(\mid\kappa\mid+\mid\gamma\mid)\chi_++q\chi_-, 
&\Phi_-=-q\chi_++(\mid\kappa\mid+\mid\gamma\mid)\chi_-.
\end{array}
\label{phiPhi}
\end{equation}

Expressing the $\chi$'s from the upper two equations in terms of the
$x$-eigenspinors $\varphi$ yield the z-eigenspinors
\begin{equation}
\chi_+={1\over2\mid\gamma\mid}\Big(\varphi_+ 
+{q\over{\mid\gamma\mid+\mid\kappa\mid}}\ \varphi_-\Big),
\qquad
\chi_-={1\over2\mid\gamma\mid}\Big(- 
{q\over{\mid\gamma\mid+\mid\kappa\mid}} 
\ \varphi_++
\varphi_-\Big),  
\end{equation}
which {\it do} indeed satisfy (\ref{dyonz}). 
For $j=q-1/2$, $\chi_-$ is missing and
$\chi_+$ is proportional to the lowest $\varphi_+^0$  in 
(\ref{dyonlowestangfunct}).  

Eliminating the $\chi'$s allows to deduce the
$y$-eigenspinors $\Phi$ from the $x$-eigenspinors $\varphi$ according to  
\begin{equation}
\Phi_+={1\over\mid\gamma\mid}
\Big(\mid\kappa\mid\varphi_++q\varphi_-\Big)
\qquad\hbox{and}\qquad
\Phi_-={1\over\mid\gamma\mid}
\Big(-q\varphi_++\mid\kappa\mid\varphi_-\Big) 
\end{equation}
which, by construction, satisfy 
\begin{eqnarray}
\vec{J}^2\Phi_{\pm}&=&j(j+1)\Phi_{\pm},
\\[4pt]
J_3\Phi_{\pm}&=&\mu\Phi_{\pm},\quad
\mu=-j,\ldots, j,
\\[4pt]
y\ \Phi_{\pm}&=&\mp\mid\gamma\mid\Phi_{\pm}.
\end{eqnarray}
Finally, $w=\vec{\sigma}\cdot\vec{r}/r$ interchanges
the $x$ and $y$ eigenspinors, 
\begin{equation}
w\,\varphi_{\pm}^{\mu}=\Phi^{\mu}_{\mp}.
\end{equation}

In contrast to what happens in the `lower' (i.e. $\rho_3=-1$) sector,
in the `upper' (i.e. $\rho_3=1$) sector 
$$
y(y-1)=\vec{L}^2-2\vec\sigma\cdot\frac{\vec{r}}{r}
$$ 
is {\it not} the square of an angular momentum 
and hence we {\it do} have $L(\gamma)=q-1$ states:  
$\mid\gamma\mid=q$, 
$\kappa=0$
for the lowest value of total angular momentum, $j=q-1/2$, and for 
$\gamma=-q$ eqn. (\ref{phiPhi}) yields (\ref{dyonlowestangfunct}), 
\begin{equation} 
\Phi_0\, \, (=\Phi_-)=\varphi_+^0,
\end{equation}
while the entire $\Phi_+$ -tower is missing.
This is a $(-1)$ eigenstate of $w$,
\begin{equation}
w\ \Phi_0=-\Phi_0.
\end{equation}

Since $\varphi_+^0$ is a $(-q)$ eigenstate of $x$, 
$\Phi_0$ is an
eigenstate of $y=x+2qw$ with eigenvalue $(+q)$. Since
\begin{equation}
\Gamma (\Gamma+1)\ \Phi^{\mu}_{\gamma}= 
L(\gamma)(L(\gamma)+1)\ \Phi^{\mu}_{\gamma}\ , 
\qquad
\Gamma (\Gamma+1)\ \varphi^{\mu}_{\gamma}
= 
L(\gamma)(L(\gamma)+1)\ \varphi^{\mu}_{\gamma},
\end{equation} 
by construction, for $j \geq q+1/2$ the eigenfunctions of $\D^2$ are found as 
\begin{equation}
\left.\begin{array}{lll}
\Psi_{\pm\mid\gamma\mid}&=u_{\pm} 
\left(\begin{array}{ll}\Phi_{\pm}\cr0\end{array}\right) 
\quad &\hbox{for} \ \ a\rho_3=1,
\\
\psi_{\pm\mid\gamma\mid} &=u_{\pm} 
\left(\begin{array}{ll}0 \cr \varphi_{\pm}\end{array}\right)  
\quad &\hbox{for} \ \ \rho_3=-1\cr 
\end{array}\right\} 
\qquad \hbox{if} \quad j \geq q+1/2, 
\end{equation}
where the radial functions $u_{\pm}(r)$ solve the non-relativistic
Coulomb-type equations 
\begin{equation}
\Big[-(\partial_r+{1\over r})^2+{L(\gamma)(L(\gamma)+1)\over r^2}
 -{2q^2 \over r}+q^2\Big]u_{\pm}
 = E^2 \ u_{\pm}.  
 \label{dyonCoulomb}
\end{equation}

By (\ref{dyonGGplus}), these are just the upper (resp. lower)
equations of
\begin{equation}
-\left(\p_r+\frac{1}{r}\right)^2-\frac{2q^2}{r}+q^2+
\frac{1}{r^2}\left(\begin{array}{cc}
(j-\half)(j+\half)&
\\[6pt]
&
(j+\half)(j+\smallover3/2)
\end{array}\right)
\end{equation}
and hence
\begin{equation}
u_{\pm}(r) \, \propto\, r^{L(\gamma)}e^{ikr}\
F\Big(L(\gamma)+1-i{q^2\over k}, 2L(\gamma)+2,-2ikr\Big),
\label{dyonaradfunct}
\end{equation}
where $k=\sqrt{E^2-q^2}$.

 For $j=q-1/2$ we get the $(2q)$ spinors  
\begin{equation}
\psi_+=u_+\left(\begin{array}{ll}0\cr\varphi_+^0\end{array}\right), 
\qquad
\hbox{sign}\ \gamma=+1
\end{equation}
in the $\rho_3=-1$ sector with $L(\gamma)=q$\footnote{$L(\gamma)=q$-values
arise in the $\rho_3=1$ sector for $\gamma=- (q + 1)$.}, with $u_+$ 
still as in (\ref{dyonaradfunct}). 

The energy levels are obtained from the poles of $F$,
$$
L(\gamma)+1-iq^2/k=-n,\; n = 0, 1, \ldots.
$$
Introducing the principal
quantum number $p=L(\gamma)+1+n\geq q+1$ we conclude that, in both
$\rho_3$ sectors,    
\begin{equation}
E_p=q^2 \Big(1- ({q\over p})^2\Big),
\qquad p=q+1,\dots
\end{equation}

The same energy is obtained if $L+n=L'+n'$. The degeneracy of a
$p\geq q+1$-level is hence $2 (p^2-q^2)$. 

If $j=q-1/2$, $(2q)$ extra states arise in the $\rho_3=1$ sector for 
$\gamma=-q$, 
\begin{equation}
\Psi_0 = u_0 
\left(\begin{array}{ll}\Phi_{0}\cr0\end{array}\right) 
\qquad \hbox{for} \ \ \rho_3 = 1 
\quad \hbox{and} \quad
\gamma=-q,
\end{equation}
where $u_0$ solves (\ref{dyonCoulomb}) with $L(\gamma)=q-1$.
The principal quantum number is now $p=q$, yielding the $2q$-fold
degenerate $0$ - energy ground states. Since $F(0,a,z) = 1$, and the lowest
$k$-value is $iq$, $u_0$ is simply
\begin{equation}
u_0=r^{q-1}e^{-q r},
\end{equation}
cf. \cite{FHO,DVdyon}. The situation is shown in Figure 3-4~:

\begin{figure}[htbp]
\centering
\includegraphics[scale=0.8]{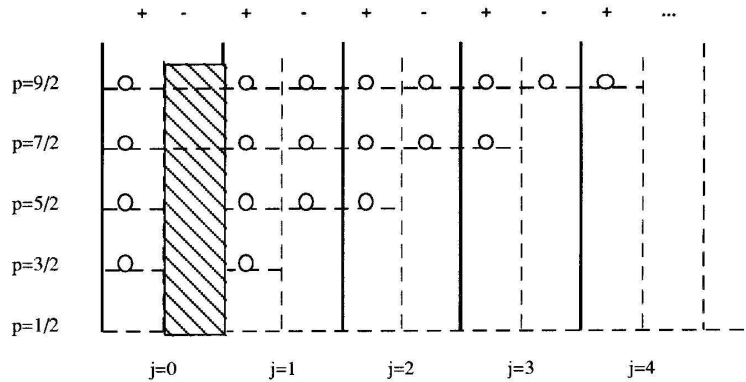}
\caption{\it The dyon spectrum in the $g=0$ sector. The sign refers to that
of $(-x)$. Each $j\geq q+1/2$ sector is doubly degenerate. For $j=q-1/2$
there are no $(-x)=-q$ states. The energy only depends on the principal
quantum number $ =L(\gamma)+1+n$.}
\label{Bloore1}
\end{figure}
\begin{figure}[htbp]
\centering
\includegraphics[scale=.8]{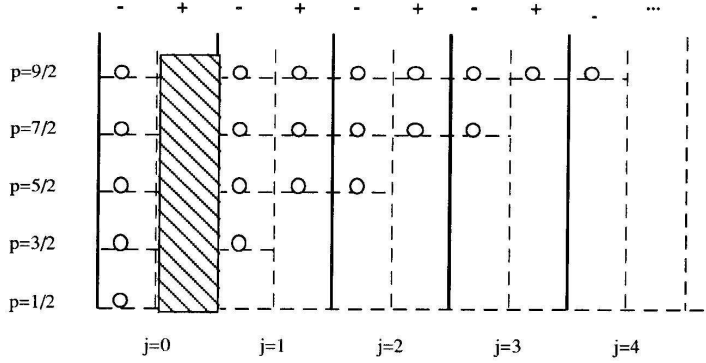}
\caption{\it The dyon spectrum in the $g=4$ sector. The sign refers to that 
of $(-y)$. Each $j \geq q + 1/2$ sector is doubly degenerate. For $j=q-1/2$
there are no $(-y)=+q$ states but $E=0$ ground states arise for 
$(-y)=-q$.}
\label{Bloore2}
\end{figure}

%%%%%%%%%%%%%%%%%%
\section{Further applications}
%%%%%%%%%%%%%%%%%%

As yet another illustration, we consider a spin $\2$
particle described by the four-component Hamiltonian
\begin{equation}
H =\left(\begin{array}{ll}
H_1&\cr &H_0
\end{array}\right)=
\2\left\{\pib^2-q{\sigmab.\hat{\bf r}\over r^2}+{\lambda^2\over r^2}
-\lambda\gamma^5{\sigmab.\hat{\bf r}\over r^2}\right\}
\label{DV85Ham}
\end{equation}
where $\lambda$ is a real constant \cite{DV85}.
The Hamiltonian (\ref{DV85Ham}) can again
be viewed as associated to a static gauge field on $\IR^4$,
\begin{equation}
{\bf A}=q{\bf A}_D,\qquad
A_4=\lambda/r,
\end{equation}
cf. (\ref{dyongaugefield}). The square of the
associated Dirac operator
\begin{equation}
\D=\left(\begin{array}{ll}
&Q^{\dagger}\cr Q&\end{array}\right) = 
\left(\begin{array}{ll}&\sigmab.\pib - i{\lambda\over r}\\
\sigmab.\pib + i{\lambda\over r} &
\end{array}\right),
\label{DV85Dsq}
\end{equation}
is precisely (\ref{DV85Ham}). 
The partner hamiltonians of the
chiral-supersymmetric Dirac operator
have again the same spectra. 

Much of the theory developed before in Sections \ref{CM} and \ref{dyon}
apply. The conserved total angular momentum is 
 (\ref{MTotangmom}) and Dirac's ${\cal K}$ is again (\ref{dyonK}).
The supercharges $Q$ and
$Q^{\dagger}$ can now be written as 
\begin{equation}
\begin{array}{llll} 
Q = &-iw\Big(\partial_r + {\displaystyle 1\over\displaystyle r} 
- {\displaystyle y\over\displaystyle r}\Big) &=
&-i\Big(\partial_r + {\displaystyle 1\over\displaystyle r} 
+ {\displaystyle x\over\displaystyle r}\Big) w, 
\\[12pt]
Q^{\dagger} = &-iw\Big(\partial_r+{\displaystyle 1\over\displaystyle r} 
-\displaystyle{x\over r}\Big)
&= &-i\Big(\partial_r + {\displaystyle 1\over\displaystyle r} 
+ {\displaystyle y\over\displaystyle r}\Big)w,
\end{array} 
\end{equation}
where 
\begin{equation}
x=z-\lambda w
\qquad\hbox{and}\qquad
y=z+\lambda w.
\end{equation}
The  Biedenharn operator,  conserved for the
quadratic dynamics, is now
\begin{equation}
\Gamma = - ({\bf\sigmab.\ell}+1+\gamma^5\lambda
w) 
\quad\hbox{i.e.}\quad
 - (z+\gamma^5\lambda w)
 \equiv -\left(\begin{array}{ll}y&\\&x  \end{array}\right).
\end{equation}
Since
$
\Gamma^2= z^2 +\lambda^2={\bf J}^2+1/4+\lambda^2-q^2,
$
the eigenvalues of $\Gamma$, 
\begin{equation}
\gamma=\pm\sqrt{(j+1/2)^2+\lambda^2-q^2},
\qquad
\hbox{sign}\ \gamma =\hbox{sign}\ \kappa,
\end{equation}
are in general again irrational.
In terms of $\Gamma$, $\D^2$ is written 
\begin{equation}
\D^2 = 
\left(\begin{array}{ll}Q^{\dagger} Q 
\cr &Q Q^{\dagger}\end{array}\right)
= - (\partial _r + {1\over r})^2 + 
{\Gamma(\Gamma +1)\over r^2}.
\end{equation}

\kikezd{The explicit solution.}

The operator $\Gamma$ can be diagonalized as in Section \ref{dyon},
cf. \cite{KYG,KY}. We get  $2$-spinors which diagonalize $z$ are
\begin{equation}
\begin{array}{llcccl}
\chi_+&=\displaystyle{1\over 2j+1}\,\Big(&\varphi_+&+ &
\displaystyle{q\over{j+1/2 +\mid\kappa\mid}}
\ \varphi_- &\Big)
\\[18pt]
\chi_-&=\displaystyle{1\over 2j+1}\,\Big(&- 
\displaystyle{q\over {j+1/2+\mid\kappa\mid}} 
\ \varphi_+&+ &\varphi_- &\Big)
\end{array}
\label{DV85chi}  
\end{equation}
where the $\phi_\pm$ are given in (\ref{dyonangfunct}).
Hence
\begin{equation}
\begin{array}{ll}
\phi_+=(\vert\kappa\vert+j+{1\over 2})\ \chi_+-\lambda\ \chi_-,
&\quad\phi_-=\,\lambda\ \chi_+ +(\vert\kappa\vert+j+{1\over 2})\chi_-
\\[16pt]
\Phi_+=(\vert\kappa\vert+j+{1\over 2})\ \chi_++\lambda\ \chi_-,
&\quad\Phi_-=-\lambda\ \chi_+ +(\vert\kappa\vert+j+{1\over 2})\chi_-
\end{array}
\end{equation}
diagonalize $x$ and $y$,
\begin{equation}
x\phi_{\pm}^{\mu} =\mp\mid\gamma\mid\phi^{\mu}_{\pm}
\qquad\hbox{and}\qquad
y\Phi_{\pm}^{\mu} =\mp\mid\gamma\mid\Phi^{\mu}_{\pm}.
\end{equation}

The operator $w =\sigmab.{\hat{\bf r}}$ interchanges
the $x$ and $y$ eigenspinors, 
\begin{equation}
w\,\phi_{\pm}^{\mu}=\Phi^{\mu}_{\mp}.
\end{equation}

For $j=q-1/2$, no $\varphi_-$ is available and $\chi_-$ is hence missing.
$\chi_+$ is proportional to the lowest $\varphi_+$ in 
(\ref{dyonlowestangfunct}). There are no 
$\phi_-$-states in the $\gamma^5 = -1$ sector and no $\Phi_+$ states
in the $\gamma_5 = 1$ sector. However, in each $\gamma^5$ sector, 
(\ref{DV85chi}) 
yields $(2q)$ $(+1)$-eigenstates of $w$, namely
\begin{equation}
(\phi_+^0)^\mu=(\Phi_-^0)^\mu
\propto\sqrt{{q+1/2+\mu\over 2q+1}}\, 
Y^{\mu-1/2}_q\left(\begin{array}{c}1\\ 0\end{array}\right) 
+
\sqrt{{q+1/2+\mu\over2q+1}}\, 
Y^{\mu+1/2}_q\left(\begin{array}{c}0\\ 1\end{array}\right).
\end{equation} 

The eigenfunctions of $\D^2$ are then found as 
\begin{equation}
\begin{array}{llll}
\left\{
\begin{array}{ll}
\Psi_{\pm\mid\gamma\mid} &=u_{\pm} 
\left(\begin{array}{c}\Phi_{\pm}\cr 0\end{array}\right) 
\\[6pt] 
\psi_{\pm\mid\gamma\mid} &= u_{\pm} 
\left(\begin{array}{c}
0\cr\phi_{\pm}
\end{array}\right) 
\end{array}
\right.
\qquad&\hbox{for}\quad
&\left\{
\begin{array}{l}\gamma^5 = 1
\\[6pt]
\gamma^5 = - 1\end{array}
\right.\quad
&\hbox{for}\;j\geq q+1/2
\\[30pt]
\left\{
\begin{array}{ll}
\Psi_-^0 &= u_-^0 
\left(\begin{array}{c}
\Phi_-\cr 0\end{array}\right) 
\\[6pt] 
\psi_+^0 &= u_+^0
\left(\begin{array}{c}
0\cr\phi_+
\end{array}\right)
\end{array}\right.
\qquad&\hbox{for}\qquad
&\left\{\begin{array}{l}\gamma^5 = 1
\\[6pt]
\gamma^5 = - 1\end{array}\right.
&\hbox{for}\;j= q-1/2
 \end{array}
\end{equation}

Thus, the radial functions $u_{\pm}(r)$ solve 
\begin{equation}
\Big[-(\partial_r +{1\over r})^2 +{\gamma
(\gamma +1)\over r^2}-2E\Big]\ u_{\pm} = 0.  
\end{equation}

This is the wave equation for a free particle except for 
the fractional \lq angular momentum' $\gamma$. Its solutions is 
hence given by the Bessel functions,
\begin{equation}
u_{\pm}(r)\,\propto\,r^{-1/2}\
J_{\vert\gamma\vert\mp\2}(\sqrt{2E}\ r).
\end{equation}

$\bullet$ For $\lambda = 0$ we recover the formulae in \cite{newsusy}.
 The well-known 
self-adjointness problem in the $j=q-1/2$ sector shows up in that
 the eigenvalue $\gamma$ vanishes in this case. 
(Self-adjointness of $\D^2$ requires in fact $\vert\lambda\vert\geq 3/2$ 
\cite{DV85}.

$\bullet$ Another interesting particular value is $\lambda=\pm q$, 
when the Biedenharn-Temple operator has half-integer eigenvalues,
\begin{equation}
\gamma =\pm (j+\2).
\end{equation}

In this case, $\gamma(\gamma+1)$ is the same for
 $-\vert\gamma\vert$ as for $\vert\gamma\vert-1$,
leading to identical solutions. 
Thus, the corresponding energy levels
are two-fold degenerate. (This only happens for  
$\vert\gamma\vert\geq\vert\gamma\vert_{min}+1$ i.e. for
$j\geq q+1/2$).
This can also be understood by noting that, for
$\lambda =\pm q$, the spin dependence drops out in 
one of the $\gamma^5$-sectors. 
For $\lambda=q$, e.g., the Hamiltonian 
(\ref{DV85Ham}) reduces to
\begin{equation}
H =\left(\begin{array}{ll}H_1&\\ &H_0\end{array}\right) = 
\2\,\left(\begin{array}{ll}
\pib^2+\displaystyle{q^2\over r^2}
-2q\displaystyle{\sigmab.\br\over r^3}&
\\[6pt]
&\pib^2+\displaystyle{q^2\over r^2}
\end{array}\right),
\end{equation}
i.e.,  $H_0$ describes a spin $0$ particle, while
$H_1=H_0-2q\sigmab.\br/r^3$ corresponds to a particle with anomalous gyromagnetic ratio $4$, cf. dyons in Section \ref{dyon}. 
The system admits hence an extra $o(3)$ symmetry, 
generated by the spin vectors
\begin{equation}
\begin{array}{ll}
{\bf S}_0= \2\sigmab
&\hbox{for} \;H_0
\\[10pt]
{\bf S}_1=U^{\dagger}{\bf S}_0U\qquad
&\hbox{for} \;H_1
\end{array},
\end{equation}
where
$
U = Q/\sqrt{H_1}$ 
and  
$U^{-1} = U^{\dagger} = 
1/\sqrt{H_1}\ Q^{\dagger}
$
are the unitary transformations which intertwine the non-zero-energy parts of 
the chiral sectors. 

Each of the partner Hamiltonians $H_1$ and $H_0$ in (\ref{DV85Ham}) have a 
 non-relativistic conformal $o(2,1)$ symmetry [7] which
combines, with $\D$ and $-i\gamma^5\D$, 
into an $osp(1/2)$ superalgebra \cite{DV85}. 

The symmetries of the problem are studied in detail \cite{DV85,HMfvH}.

\kikezd{Acknowledgment}
I am indebted to Roman Jackiw for interesting
correspondence.

%%%%%%%%%%%%%%%%%%%%%%%%%%%%%%%%%%%%%%%%%%%%%%%%%%%%%%%%%%%%%%%%%%%%%%%%

\end{document}